\documentclass{aa}
\usepackage{txfonts,rotating}
\usepackage{graphicx}
\begin{document}

\def\msun{\hbox{M$_\odot$}}

\title{Extended Main Sequence Turn-Offs in Low Mass Intermediate Age Clusters}

\author{Andr\'es E. Piatti\inst{1,2}
\and 
Nate Bastian\inst{3}
}

\institute{Observatorio Astron\'omico, Universidad Nacional de C\'ordoba, Laprida 854, 5000, 
C\'ordoba, Argentina;\\
\email{andres@oac.uncor.edu}
\and
Consejo Nacional de Investigaciones Cient\'{\i}ficas y T\'ecnicas, Av. Rivadavia 1917, C1033AAJ,
Buenos Aires, Argentina 
\and
Astrophysics Research Institute, Liverpool John Moores University, 146 Brownlow Hill, Liverpool L3 5RF, UK
}

\date{Received / Accepted}

\abstract{We present an imaging analysis of four low mass stellar clusters ($\lesssim5000$~\msun)
in the outer regions of the LMC in order to shed light on the extended main sequence turn-off
 (eMSTO) phenomenon observed in high mass clusters.  The four clusters have ages between
 $1-2$~Gyr and two of them appear to host eMTSOs.
The discovery of eMSTOs in such low mass clusters  - $>$ 5 times less massive than the eMSTO clusters previously studied - suggests that mass is not the controlling
 factor in whether clusters host eMSTOs.  Additionally, the narrow extent of the eMSTO in the two older
 ($\sim2$~Gyr) clusters is in agreement with predictions of the stellar rotation scenario,
 as lower mass stars are expected to be magnetically braked, meaning that their CMDs should 
be better reproduced by canonical simple stellar populations.  We also performed a structural
 analysis on all the clusters and found that a large core radius is not a requisite
for a cluster to exhibit an eMSTO.}


\titlerunning{eMSTOs in low mass clusters}
\authorrunning{A.E. Piatti and N. Bastian}

\maketitle

\markboth{A.E. Piatti and N. Bastian: eMSTOs in low mass clusters }{}

\section{Introduction}

The Large Magellanic Cloud (LMC) hosts a relatively large cluster population, with significant 
numbers of young (less than a few hundred Myr) and intermediate age ($\sim1-3$~Gyr) clusters. 
 Due to the relative low extinction towards most of them, and the lack of a significant confusion along the
 line-of-sight (and known distances), these clusters can often be studied in more detail than 
their Galactic counterparts.  As such, their stellar populations allow for much more stringent 
tests on isochrones and evolutionary tracks, and the high mass clusters provide an opportunity to
 study relatively short stellar evolutionary phases.

One surprise that has come from studying the stellar populations of the LMC (and of its companion,
the SMC) clusters is 
that many of the young and intermediate age clusters show main sequence turn-offs that are broader 
than expected for a single stellar population (SSP; including photometric errors and the effects of 
binaries).  This phenomenon was originally discovered in intermediate age clusters
 \citep[e.g.][]{bertellietal2003,mb07} but has recently also been found in  young massive clusters \citep[e.g.][]{miloneetal15,correntietal15,miloneetal2016,bastianetal2016}.
It is now clear that the extended main sequence turn-off (eMSTO) phenomenon is a
common feature of massive young  \citep{bsv13,niederhoferetal15b,miloneetal2016,bastianetal2016} and intermediate age clusters  \citep[e.g.][]{mietal09,goudfrooijetal2011,petal14}.
One interpretation of the eMSTOs 
is that clusters host an age spread of $100-700$~Myr within them  \cite[e.g.][]{mietal09,goudfrooijetal14}.
\citet{goudfrooijetal14} suggested a relation between the inferred age spread within the clusters 
and the escape velocity of the clusters, when the clusters were young.  In order to correct the 
current escape velocities of the clusters to the near-initial values, they
adopted a model of extreme mass loss, meaning that the clusters would have been significantly 
more massive in the past.  Based on these inferred initial masses (and also that the clusters were significantly more compact in the past) the authors suggest a minimum escape velocity of $\sim15$~km/s, above which clusters are able to retain stellar ejecta (and accrete material from the surroundings) and form subsequent generations of stars.

\citet{niederhoferetal15b} and \citet{niederhoferetal15c}, however, found a  stronger relation between the inferred age 
spread and the age of the cluster, with younger clusters hosting smaller age spreads, and older
 clusters hosting larger spreads.  This agrees with the observation that younger ($<500$~Myr) 
clusters in the LMC do not show age spreads as large as $300-700$~Myr, which are commonly 
inferred for the intermediate age clusters \citep[e.g.][]{bsv13,niederhoferetal15}.  \citet{niederhoferetal15b} have shown that younger ($<1$~Gyr) massive clusters also show evidence of age spreads that continue the age-spread - age relation down to ages of  $\sim50$~Myr.    Such a correlation between the inferred age spread and the cluster age suggests
 that the eMSTO phenomenon is not due to true age spreads, but rather to stellar 
evolutionary effects. 

% This is also supported by the fact that the post main sequence features in many of  the colour-magnitude diagrams (CMDs) of clusters that host eMSTOs do not show evidence for such large age spreads \citep{lietal14,bn15,niederhoferetal15c}. 

One such stellar evolutionary induced effect that has been suggested is stellar rotation \citep{bdm09}.  Rotation
 can affect the evolution of the star through increased internal mixing and can also change the 
colour/magnitude of a star due to geometrical effects of stellar flattening. 
\cite{bh15b} used the rotating isochrones from the Geneva models \citep{ekstrometal12,georgyetal14}
and showed that indeed stellar rotation can reproduce the eMSTOs observed
 in intermediate age clusters. \citet{niederhoferetal15b} used the same models and found that they 
predicted a relation between an apparent age spread and cluster age, that matches the observations
 well. \cite{dantonaetal15} modelled the CMD of NGC~1856 ($\sim300$~Myr, $\sim10^5~\msun$) 
and found that rotation could explain the observed eMSTO and also an observed splitting in the CMD 
just below the MSTO. Whether or not rotation can match the observed ‘dual main sequences’ in young clusters \citep[e.g., NGC\,1856][]{miloneetal15} is still an open question.  In principle, rotation is a possible explanation, however the required rotational velocity distribution may not be able to adequately reproduce the MSTO 
\citep[e.g.][]{miloneetal2016}.  Further effort in the modelling of stellar rotation is required to address this.

However, the observed splitting of the main sequence
is not consistent with an age spread. Future HST based imaging surveys of more young clusters will be able to explicitly test the prediction of the rotation scenario, as the $\sim100$~Myr clusters should show extended main sequence turn-offs with inferred ages 
spreads of $\sim30$~Myr.

As discussed in \citet[][see also \cite{bdm09}]{bh15b}, if rotation is the
 underlying cause of the eMSTO phenomenon, we would expect a peak in the inferred age spread at 
around $1-1.5$~Gyr, followed by a decrease to older ages.  For lower mass stars ($\lesssim1.7$~\msun)
 magnetic breaking becomes a dominant effect and is expected to slow down rotating stars, minimising the
 effect of rotation on the host star's properties.  Such a turn-over in the $\Delta$(age)-age plot has 
been reported by \citet{niederhoferetal15c}. 

While the eMSTO phenomenon has now been routinely observed in the CMDs of massive clusters 
(with ages between $\sim100$~Myr and $\sim2$~Gyr)  there have been suggestions that such a feature is
 missing in low mass clusters \citep[e.g.][]{goudfrooijetal14}.  If age spreads were the cause, 
this could be due to the lower gravitational potential of these clusters, which would not allow them 
to retain or accrete the gas required to form a second generation of stars.  Hence, a clear 
prediction of this scenario is that low mass ($1000-5000$~\msun) clusters should not show the eMSTO 
phenomenon \citep{cs11,goudfrooijetal14}.  If stellar rotation was the cause, the lack of eMSTOs might be due to the lower number 
of stars on the MSTO, making any extension difficult to see, or possibly by a link between the 
stellar rotation distribution and cluster mass.  However, low mass clusters might show the eMSTO in 
the case of stellar rotation.

If age spreads (of order $\sim300-700$~Myr) were common in massive clusters, we should see evidence of 
ongoing star-formation (or extended SFHs) in either the observed CMDs of young massive clusters and 
in the integrated spectra of the clusters.  This has been searched for in clusters with ages between
 a few 10s of Myr and $\sim600$~Myr and masses between $10^4 - 10^7$~\msun.  No evidence for such
 extended star-formation episodes has been found 
\citep{bastianetal13,cabrerazirietal14, cabrerazirietal15, niederhoferetal15b}.  
This provides evidence against the idea that the eMSTOs are caused by age spreads.

We present an analysis of four low mass ($\lesssim5\times10^3$~\msun) clusters in the outer regions of the LMC 
with ages between 1 and 2~Gyr in order to see whether they host the eMSTO phenomenon.  Their location 
in the outer parts of the LMC means that they are not expected to have lost significant fractions of 
the initial stellar population due to cluster dissolution \citep[e.g.][]{baetal13}.  If eMSTOs
 are found this would show that cluster mass (or escape velocity) is not a controlling factor in their 
appearance, which in turn would suggest that age spreads are not present, i.e. that the eMSTOs are 
caused by another effect, such as stellar rotation.

This paper is organised as follows. Section 2 provides a description of
the data on which this study is based and the data reduction
procedures applied. We estimate the extinction, age and inferred age spread for each of the clusters in Section 3 
and their structural parameters in Section 4.  We discuss our results in Section 5 and 
present our main conclusions in Section 6.

\section{Observations and data reductions}

\begin{figure}
	\includegraphics[width=\columnwidth]{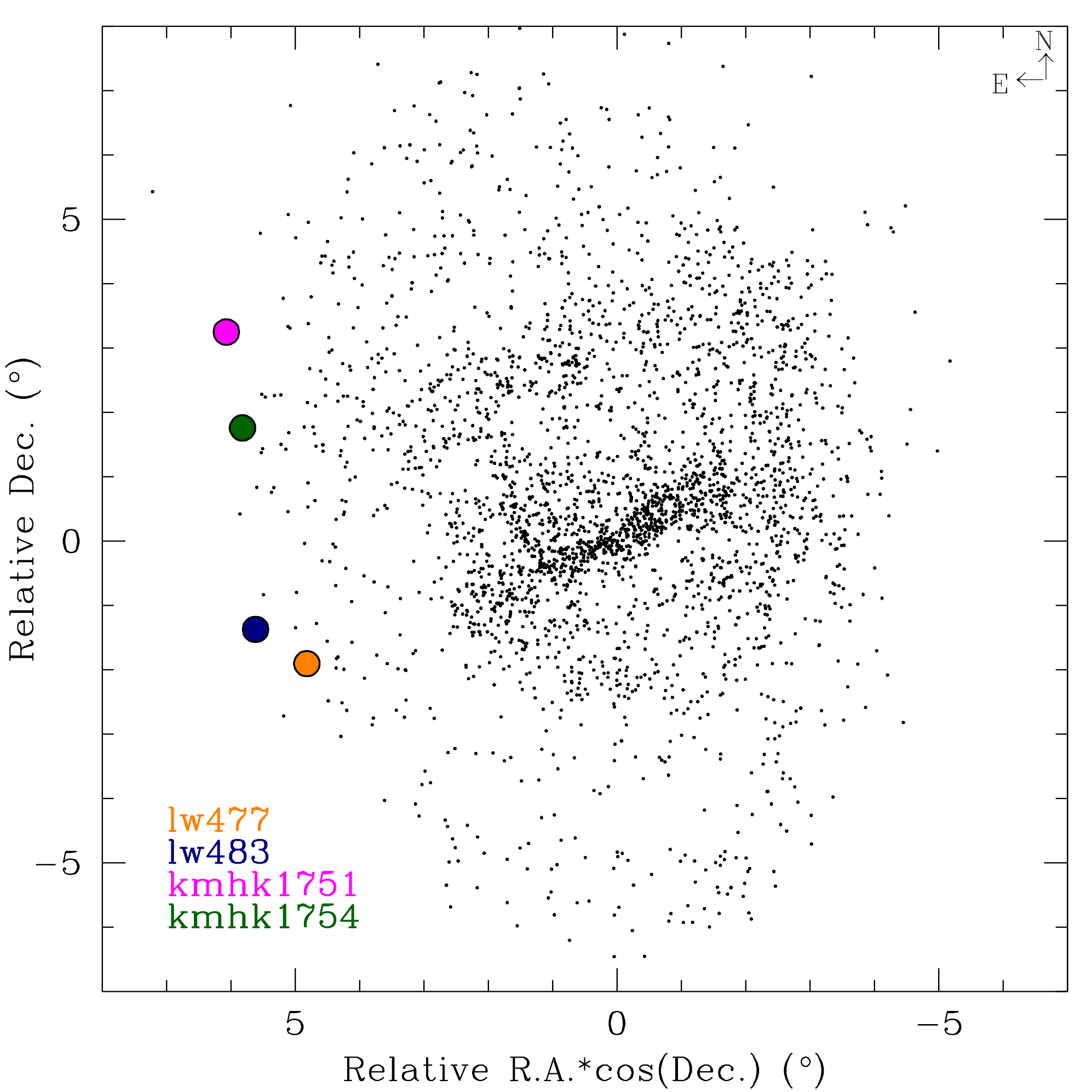}
    \caption{Spatial distribution of the \citet{betal08}'s catalogue of
  star clusters in the LMC centred at R.A. = 05$^{\rm h}$ 23$^{\rm m}$
  34$^{\rm s}$, Dec. = $-$69$\degr$ 45$\arcmin$ 22$\arcsec$ (J2000),
  projected onto the sky. The objects studied here are highlighted with
colour.}
    \label{fig:fig1}
\end{figure}

The observations presented here are part of a programme aimed at producing high quality
CMDs and radial profiles from $g,i$ photometry of unstudied LMC star clusters 
located in the LMC's outskirts and spread in two
regions of $\sim$3$^o$$\times$3$^o$ to the south-east following the direction of the Bar and at $\sim$ 
45$^o$ clockwise from that direction (see Fig.~\ref{fig:fig1}).
The selected regions were pointed to  provide an 
excellent feedback in exploring comprehensively whether tidal effects have reached the outskirts 
of the LMC. Additionally, searching for old clusters with masses smaller than 10$^4$M$_{\odot}$ which, 
according to the results 
of recent single stellar population models should be observable \citep{baetal13}, as well as
looking for multiple
stellar populations within the cluster sample are presently active issues which deserves much more
attention.
Since LMC clusters can be as old as 2-3 Gyr --in addition to the 15 known ancient globular clusters--
\citep{p11a,pg13}, we expect that our seleted clusters are at most as old as this
age regime. The Main Sequence turnoff (MSTO) of a 2-3 Gyr old cluster is at 
$V$ ($\approx$ $g$)  $\sim$ 21 mag \citet{petal14}, which corresponds to an F0 V dwarf star \citep{shk82}.

We obtained images of 4 previously selected unstudied LMC clusters with the Gemini South telescope 
and the GMOS-S instrument through $g$ and $i$ filters.  In imaging mode GMOS-S has a 
field-of-view of approximately 5.5$\arcmin$$\times$5.5$\arcmin$ at a scale of 0.16 arcsec per 
(2x2 binned) pixel.  The detector array consists of three  2K$\times$4K Hamamatsu chips 
arranged in a row. Observations were executed in 
queue mode (under programme GS-2015A-Q-44, PI: Piatti) which enabled the 
data to be obtained in excellent seeing and under photometric conditions.  
Short and long exposure images were taken in each filter ($g,i$) to provide coverage of 
bright cluster red giant branch 
stars as well as stars at least three magnitudes below the MSTO.
The log of observations is presented in Table~\ref{tab:table1}, where the main 
astrometric, photometric and observational information is summarized.
The data reduction followed the procedures documented in the Gemini Observatory 
webpage\footnote{http://www.gemini.edu}
and utilized the {\sc gemini/gmos} (v1.13) package in IRAF\footnote{IRAF is distributed by the National 
Optical Astronomy Observatories, which is operated by the Association of 
Universities for Research in Astronomy, Inc., under contract with the National 
Science Foundation.}. We performed overscan, trimming, bias subtraction, flattened all data images, etc., 
once the
calibration frames (zeros and flats) were properly combined.

\begin{table*}
\centering
\caption{Observation log of selected clusters.}
\label{tab:table1}
\begin{tabular}{lcccccc} 
\hline\hline
Star Cluster  & $\alpha_{\rm 2000}$ & $\delta_{\rm 2000}$  & 
filter & exposures & airmass & seeing  \\
     & (h m s)  & ($\degr$ $\arcmin$ $\arcsec$) & 
     &    (times $\times$ sec) &         & ($\arcsec$)\\
\hline

LW\,477 &  06 24 49 & -71 39 32 & $g$ & 1$\times$300 + 1$\times$30 & 1.46, 1.45 & 0.94, 0.96 \\
        &           &           & $i$ & 1$\times$280 + 1$\times$30 & 1.47, 1.47 & 0.80, 0.91 \\
LW\,483 &  06 33 01 & -71 07 40& $g$ & 1$\times$300 + 1$\times$30 & 1.50, 1.49 & 1.07, 0.91 \\
        &           &           & $i$ & 1$\times$280 + 1$\times$30 & 1.51, 1.51 & 0.85, 0.86 \\
KMHK\,1751 & 06 24 28  & -66 30 22 & $g$ & 1$\times$300 + 1$\times$30 & 1.56, 1.55 & 1.13, 1.21 \\
       &           &           & $i$ & 1$\times$280 + 1$\times$30 & 1.59, 1.58 & 0.99, 0.85 \\
KMHK\,1754 & 06 25 41 & -67 59 47 & $g$ & 1$\times$300 + 1$\times$30 & 1.43, 1.42 & 1.06, 1.02 \\
    &           &           & $i$ & 1$\times$280 + 1$\times$30 & 1.44, 1.44 & 0.90, 0.90 \\
\hline
\end{tabular}
\end{table*}

Around 20-50 independent magnitude measurement of stars in the standard fields E5$\_$b F3, 160100-600000 F2, 
180000-600000 F1 and 220000-595900 F2
were derived per filter using
the {\sc apphot} task within IRAF, in order to secure the transformation
from the instrumental to the $gi$ standard system.  Standard stars were
distributed over an area similar to that of the GMOS array, so that we measured magnitudes of
 standard stars in 
each of the three chips.  The relationships between
instrumental and standard magnitudes were obtained by fitting
the following equations:

\begin{equation}
g = g_1 + g_{std} + g_2\times X_g + g_3\times (g-i)_{std}
\end{equation}

\begin{equation}
i = i_1 + i_{std} + i_2\times X_i + i_3\times (g-i)_{std}
\end{equation}

\noindent where $g_j$, and $i_j$ (j=1,2,3) are the fitted coefficients, and
$X$ represents the effective airmass. 
%We adopted the mean extinction coefficient values 
%carefully measured and provided to the whole community by the Geminy Observatory staff. 
We solved the transformation equations for the three chips with the {\sc fitparams}
task in IRAF, simultaneously, and found mean colour terms of -0.025 in $g$ and -0.024 in $i$, and
extinction coefficients of 0.315 in $g$ and 0.216 in $i$;
the rms errors from
the transformation to the standard system being 0.029 mag for $g$ and 0.031 for $i$, respectively, 
indicating an excellent match to the standard system.

The stellar photometry was performed using the star-finding and point-spread-function (PSF) fitting 
routines in the {\sc daophot/allstar} suite of programs \citep{setal90}. 
For each frame, a quadratically varying 
PSF was derived by fitting $\sim$ 100 stars, once the neighbours were eliminated using a preliminary PSF
derived from the brightest, least contaminated 30-40 stars. Both groups of PSF 
stars were interactively selected. We then used the {\sc allstar} program to apply the resulting PSF to the 
identified stellar objects and to create a subtracted image which was used to find and measure magnitudes of 
additional fainter stars. This procedure was repeated three times for each frame. Finally, 
we computed aperture corrections from the comparison of PSF and aperture magnitudes by using the 
neighbour-subtracted PSF star sample. After deriving the photometry for all detected objects in
each filter, a cut was made on the basis of the parameters
returned by {\sc daophot}. Only objects with $\chi$ $<$2, photometric error less than 2$\sigma$ above 
the mean error at a given magnitude, and $|$SHARP$|$ $<$ 1.0 were kept in each filter, and then the
remaining objects in the $g$ and $i$ lists were matched with a
tolerance of 1 pixel and raw photometry obtained. 

We combined all the independent instrumental magnitudes using the stand-alone {\sc daomatch} and 
{\sc daomaster} programs\footnote{Program kindly provided by P.B. Stetson}. As a result, we produced 
one dataset per cluster containing
the $x$ and $y$ coordinates for each star and the respective robust weighted mean magnitudes.
The gathered photometric information were standardized using 
equations (1) to (2).The resulting standardized photometric 
tables
consist of a running number per star, Right Ascension and Declination (J2000.0), the mean $g$ 
magnitudes and $g-i$ colours, their respective errors $\sigma(g)$ and 
$\sigma(g-i)$, and the number of observations per star. 
%We adopted the photometric errors provided by {\sc allstar} for stars with only one measure. 
Tables 2 to 5 provide this information for LW\,477, LW\,483, KMHK\,1751 and
KMHK\,1754, respectively. Only a portion 
of Table 2 is shown here for guidance regarding their form and content. The whole content of 
Tables 2-5 is 
available in the online version of the journal. These tables were used for every subsequence analysis in this work.

\begin{table*}
\centering
\caption{CCD $gi$ data of stars in the field of LW\,477.}
\label{tab:table2}
\begin{tabular}{lccccccc}\hline\hline
Star & RA(J2000)  & DEC(J2000) & $g$ & $\sigma$($g$) & $g-i$ & $\sigma$$(g-i)$ & n \\
     & (h:m:s) & ($\degr$ $\arcmin$ $\arcsec$) & (mag) & (mag) & (mag) & (mag)  \\\hline
-    &   -     &   -     &  -    &  -    &   -   &   -     \\
     31&   06:25:22.797& -71:37:31.81  & 21.080 &   0.006  &  0.437  &  0.009  & 2\\
     32&   06:25:22.704& -71:39:15.55  & 21.336 &   0.016  &  0.312  &  0.017  & 2\\
     33&   06:25:22.676& -71:38:24.46  & 18.153 &   0.003  &  0.940  &  0.004  & 2\\
-    &   -     &   -     &  -    &  -    &   -   &   -     \\
\hline
\end{tabular}
\end{table*}

In order to evaluate the influence of the photometric 
errors, crowding effects and the detection limit on the cluster fiducial characteristics in the CMDs,
we first examined the quality of our photometry. 
To do this, we derived the completeness
level at different magnitudes by performed artificial star tests on a long exposure image per filter
 and per cluster.
We used the stand-alone {\sc addstar} program in the {\sc daophot}
package \citep{setal90} to add synthetic stars, generated bearing in mind the colour and magnitude 
distributions 
of the stars in the CMDs (mainly along the main sequence and the red giant branch), as well as the 
radial stellar density profiles 
of the cluster fields. We evaluated the effect of crowding in five different rings centred on
the clusters, between:
0--8, 8--16, 16--32, 32-64 and 64-160 arcsec, and the dependence with the magnitude in bins of 0.5 mag wide
along the whole magnitude dynamical range. 
We added a number of stars equivalent to $\sim$ 5$\%$ of the measured stars in order to avoid
significantly more crowding synthetic images than in the original images. On the other hand, to
 avoid small  number statistics, very extensive artificial star tests were performed on each image. 
We used the option of entering the number of photons
per ADU in order to properly add the Poisson noise to the star images.

\begin{figure}
	\includegraphics[width=\columnwidth]{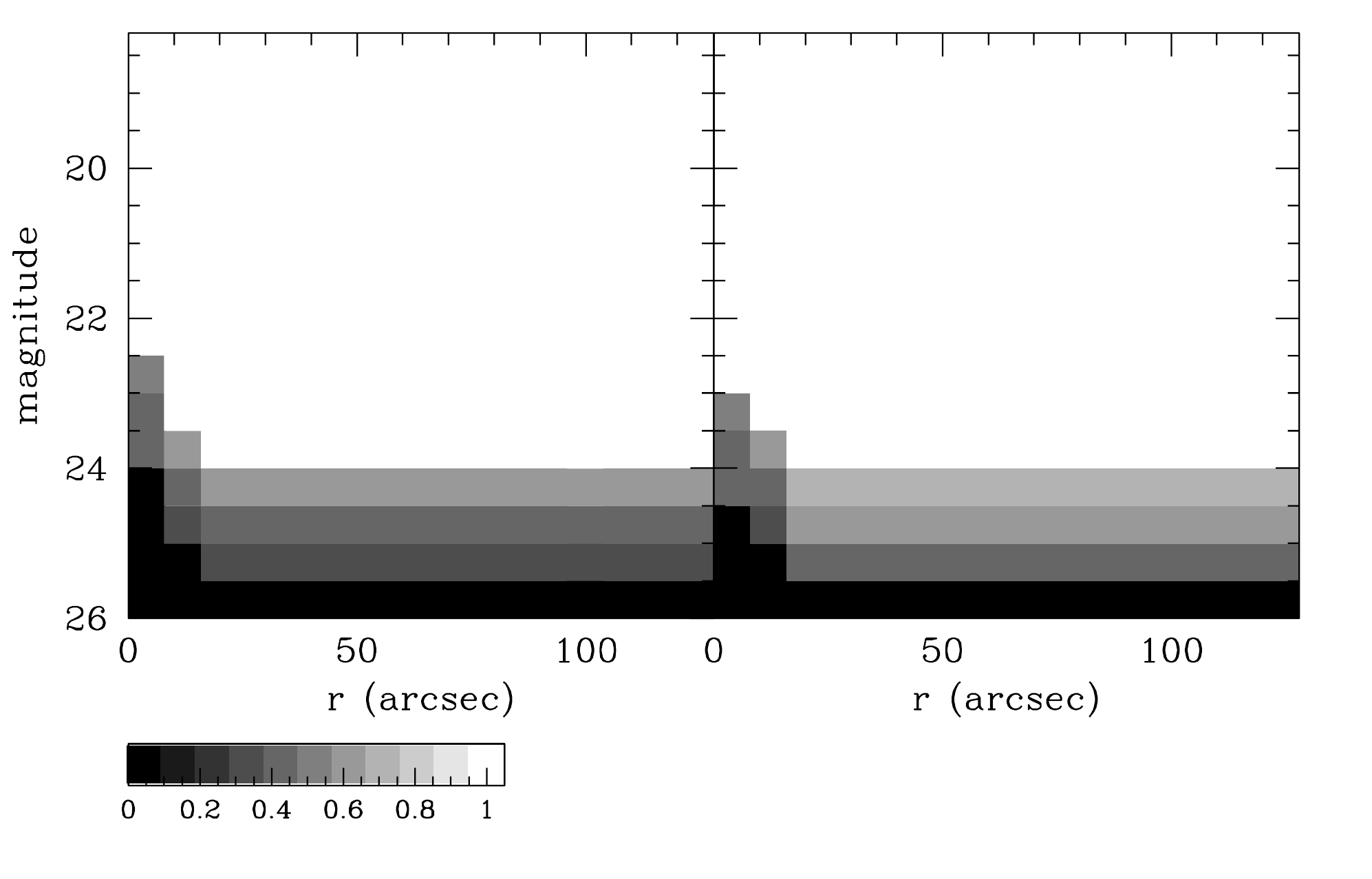}
    \caption{Completeness level in
$g$ (left) and $i$ (right) bands are shown as a function of the distance from the centre
of LW\,483.}
    \label{fig:fig2}
\end{figure}

\begin{figure*}
	\includegraphics[width=\textwidth]{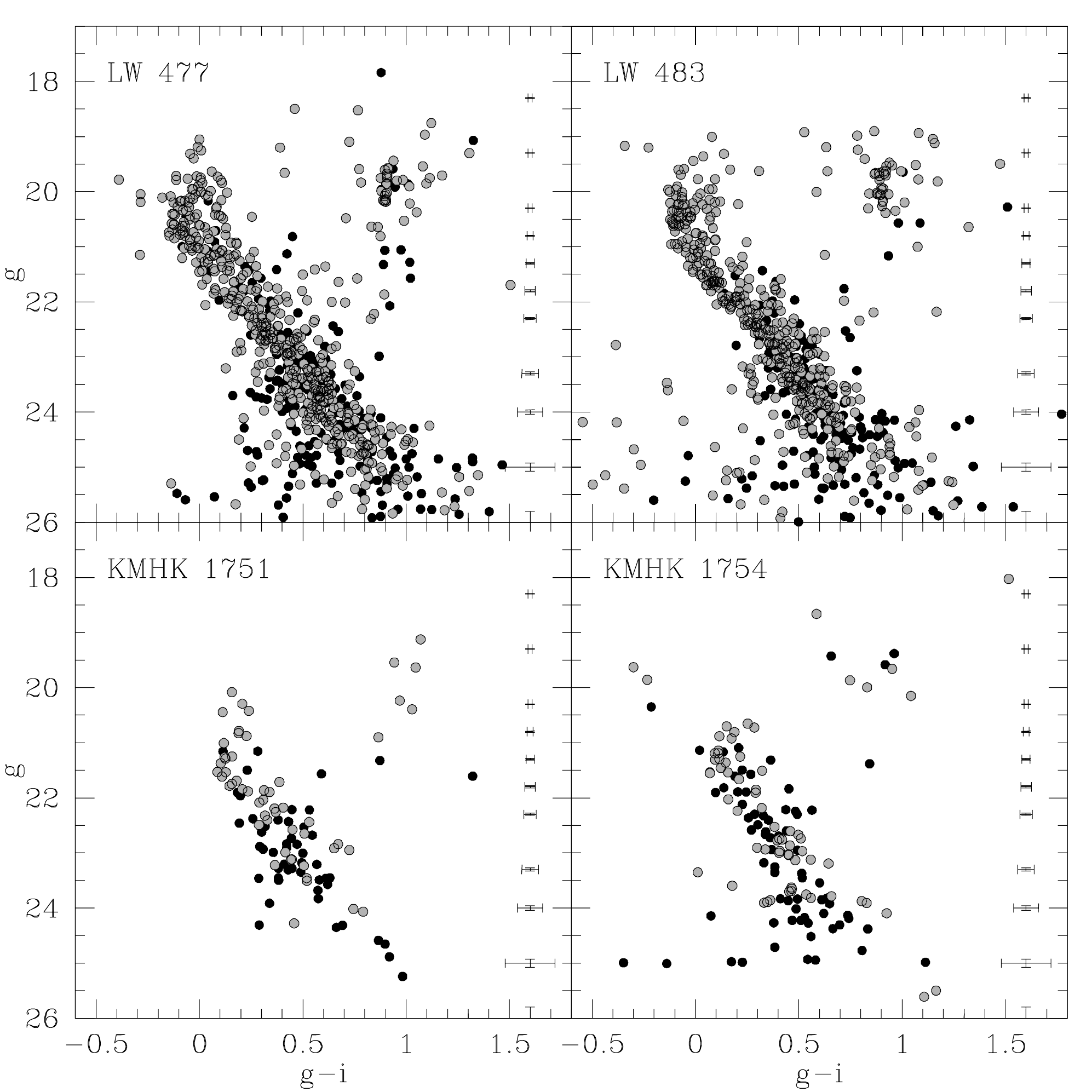}
    \caption{$g$ versus $g-i$ CMDs for the studied clusters. Black filled
circles represent stars measured inside the defined field areas, while clear grey
filled circles correspond to stars located in the cluster regions (same area size 
as field regions) which were not eliminated from the field star decontamination procedures.}
    \label{fig:fig3}
\end{figure*}

We then repeated the same steps to obtain the photometry of the synthetic images as described above, i.e., 
performing three passes with the {\sc daophot/allstar} routines, making a cut on the basis of the parameters
returned by {\sc daophot}, etc. 
The errors and star-finding efficiency was estimated by comparing the output 
and the input data for these stars - within the respective magnitude and colour bins - 
using the {\sc daomatch} and {\sc daomaster} tasks. In our experiments an artificial star
 is considered recovered if it is found 
an input–output difference in magnitude less than
0.5 mag, and at the
same time satisfying all the photometric selection criteria.
In Fig.~\ref{fig:fig2} we show the resultant completeness fractions as a function of magnitude and 
distance to the centre of LW\,483, the most massive cluster
in our sample.  
Fig.~\ref{fig:fig2} shows that the 50$\%$ completeness level is reached at $g$ $\sim$ 23.5 mag and 
$i$ $\sim$ 24.0 mag, independently of the distance from the cluster centre (outside the inner 20"). 
Thus, we conclude that our photometry is able to reach the 50\% completeness level 2-3 magnitudes below
the MSTO for the innermost cluster regions.
The behaviour of the magnitude and colour errors is represented by error bars in the CMDs shown in 
Fig.~\ref{fig:fig3}.

\section{Cluster fundamental parameters}

\subsection{Cluster ages and extinctions}

We first constructed CMDs representing
the field along the line of sight towards the
individual clusters, which we then used to clean the cluster CMDs. We
employed the cleaning procedure developed by \citet[see their
  fig. 12]{pb12}. The method compares the extracted cluster CMD
 to a field CMD composed of stars located reasonably far from the
object, but not too far so as to risk losing the local field-star
signature in terms of stellar density, luminosity function and
colour distribution. We decided to clean a circular region around the
cluster centre with a radius equal to the cluster radius ($r_{cls}$, see Sect. 4). 
The field region was designed to cover an
equal area as that of the cluster  for an annulus 
 -outer and inner radii equal to 2.236 and 2.0 times $r_{cls}$-
centred on  the cluster.

Comparisons of field and cluster CMDs have long been done by comparing
the numbers of stars counted in boxes distributed in a similar manner
throughout the CMD. However, since some parts of the CMD are more
densely populated than others, counting the numbers of stars within
boxes of a fixed size is not universally efficient. For instance, to
deal with stochastic effects at relatively bright magnitudes (e.g.,
fluctuations in the numbers of bright stars), larger boxes are
required, while populous CMD regions can be characterized using
smaller boxes. Thus, the use of boxes of different sizes distributed in
the same manner throughout both CMDs leads to a more meaningful
comparison of the numbers of stars in different CMD regions. By
starting with reasonably large boxes -- ($\Delta$($g$),$\Delta$($g-i$)) = 
(1.0, 0.5) mag -- centred on each
star in the annular field CMD and by subsequently reducing their sizes
until they reach the stars closest in magnitude and colour,
separately, we defined boxes which result in the use of larger areas in
field CMD regions containing a small number of stars, and vice versa.
Next, we plotted all these boxes
for the field CMD on the cluster CMD and subtracted the star located
inside them and closest to each box centre.

We used the cleaned cluster CMDs to estimate some fundamental cluster
parameters by matching the observations with the theoretical
isochrones of \citet{metal08}. In performing this task, one has to
deal with the reddening, distance, age and
metallicity. Our strategy consisted of obtaining the reddening values
from an independent source, assuming the mean LMC distance modulus for
all clusters and adopting for the mean cluster ages the ages of the
isochrones which best reproduced their CMD features. We started with
isochrones for a metallicity of [Fe/H] = -0.4 dex (Z = 0.006, Z$_{\odot}$ = 0.0152),
which corresponds to the mean LMC cluster metal content during the
last $\sim 2$--3 Gyr \citep{pg13}, and employed isochrones for other
metallicities as required.

 We made
use of the NASA/IPAC Extragalactic Data
base\footnote{http://ned.ipac.caltech.edu/. NED is operated by the Jet
  Propulsion Laboratory, California Institute of Technology, under
  contract with NASA.} (NED) to infer Galactic foreground reddening
values, $E(B-V)$, for our cluster list (see Table~\ref{tab:table6}). 
According to \citet{dutraetal01}, who studied spectral properties of
galaxies behind the LMC, the outermost eastern part of the LMC is optically thin,
characterized by an average combined foreground and internal $E(B-V)$ colour
excesses of 0.06$\pm$0.03 mag. The latter is in excellent agreeement with those
listed in Table~\ref{tab:table6}. As for the cluster distance moduli, we 
adopted the same distance modulus for all
clusters $(m-M)_o = 18.49 \pm 0.09$ mag \citep{dgetal14} and 
$g - M_g = (m-M)_o +  3.738 \times E(B-V)$, for $R_V = 3.1$
\citep{cetal89,getal13}, since by considering an average depth for the
LMC disc of (3.44$\pm$1.16) kpc \citep[$\Delta$($(m-M)_o$) $\sim$ 0.15 mag][]{ss09}, 
we derived a smaller age
difference than that resulting from the isochrones (characterized by
the same metallicity) bracketing the observed cluster features in the CMD.

In the matching procedure, we used subsets of isochrones ranging from
$\Delta \log(t$ yr$^{-1}) = $-0.3 to +0.3,  once they were 
properly shifted by the corresponding reddening and LMC distance 
modulus,  straddling the initial rough age estimates.
Notice that by matching different single stellar population (SSP)
isochrones we do not take into account
the effect of the unresolved  binaries  or stellar rotation   but focus on the possibility that any
unusual broadness at the MSTO might come from the presence of populations of different ages. 
\citet{maetal08,metal09,getal09,p13}, among others, showed that
a significant fraction of unresolved binaries is not enough to reproduce 
the eEMSTOs seen in their studied clusters.
%, while stellar rotation has not driven the
%whole MSTO broadness in all the cases \citep{gietal09,letal14}. 
%Consequently, the matching of SSP isochrones results overall justified. 
% removed by NB - Oct. 14th
Moreover, by closely inspecting the
matched cluster MSTO regions we have a hint for any uncommon broadness
in the studied cluster sample.

\begin{figure*}
	\includegraphics[width=\textwidth]{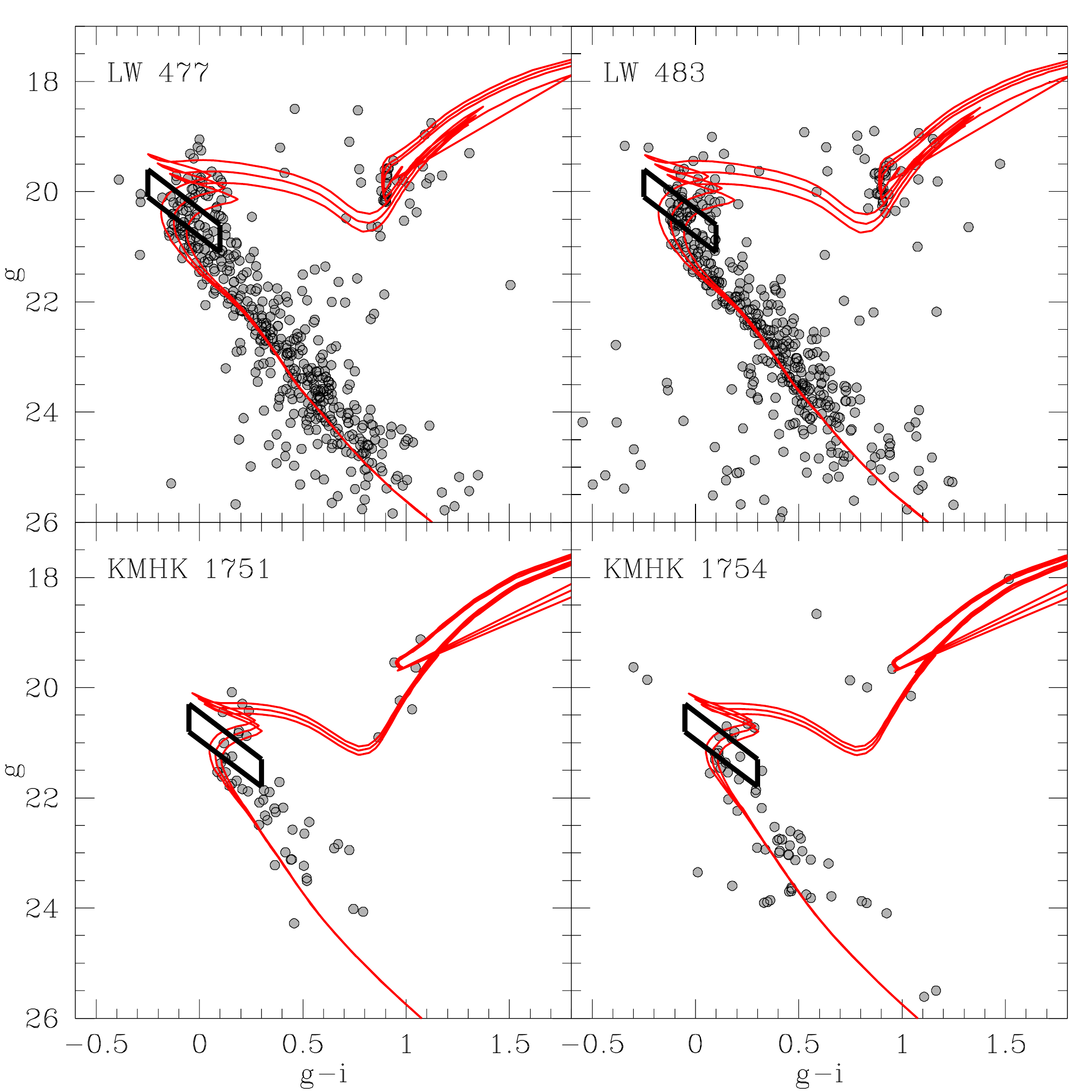}
    \caption{Cleaned $g$ versus $g-i$ CMDs for stars located within $r_{cls}$  
with theortical isochrones superimposed ([Fe/H]= -0.4 dex). The isochrones are separated
according to the errors quoted in Table~\ref{tab:table6}.}
    \label{fig:fig4}
\end{figure*}

Finally, we adopted the cluster age as the age of the isochrone which best 
reproduced the cluster's main features in the CMD, namely, the cluster's Main Sequence (MS) and
 red clumps (RCs). 
%We found that isochrones bracketing the derived mean age by the $\sigma$(age)
%values quoted in Table~\ref{tab:table6} represent the overall age uncertainties
%owing to the observed dispersion in the cluster CMDs. 
%These age uncertainties are thought to mainly represent the overall dispersion along the 
%Subgiant Branch as well as 
%the position of the RC, rather than a measure of the MSTO spread.
%Nevertheless, the adopted age uncertainties relatively reflect the observed MSTO 
%broadness. 
% removed by NB - Oct. 14th
Fig.~\ref{fig:fig4}  shows the results 
of isochrone matching along with the associated age uncertainties ( from Table~6). 
As far as we are aware, none of the four clusters had been previously studied.

\setcounter{table}{5}
\begin{table*}
\centering
\caption{Fundamental properties of LMC star clusters.}
\label{tab:table6}
\begin{tabular}{@{}lcccccc}\hline\hline
Star cluster & $E(B-V)$ & [Fe/H] &  age          &  $\Delta$(age)$^a$ & $t_r$ & $M_{cls}$ \\
             &          &  (dex) &  (Gyr)        & (Myr)      &   (Myr) & (10$^3$~M$_{\odot}$)  \\\hline
LW\,477      & 0.070    & -0.4   & 1.25$\pm$0.15 & 310$\pm$30 &  225    & 2.2$\pm$1.5 \\ 
LW\,483      & 0.075    & -0.4   & 1.25$\pm$0.15 & 300$\pm$30 &  310    & 5.5$\pm$3.8\\
KMHK\,1751   & 0.055    & -0.4   & 2.00$\pm$0.10 & $<$190$\pm$20 &  150    & 1.2$\pm$0.8\\
KMHK\,1754   & 0.055    & -0.4   & 2.00$\pm$0.10 & $<$180$\pm$20 &  160    & 1.4$\pm$0.9\\
\hline
\end{tabular}

\noindent $^a$ $\Delta$(age)= $FWHM_{MSTO}$

\end{table*}

\begin{figure*}
	\includegraphics[width=\textwidth]{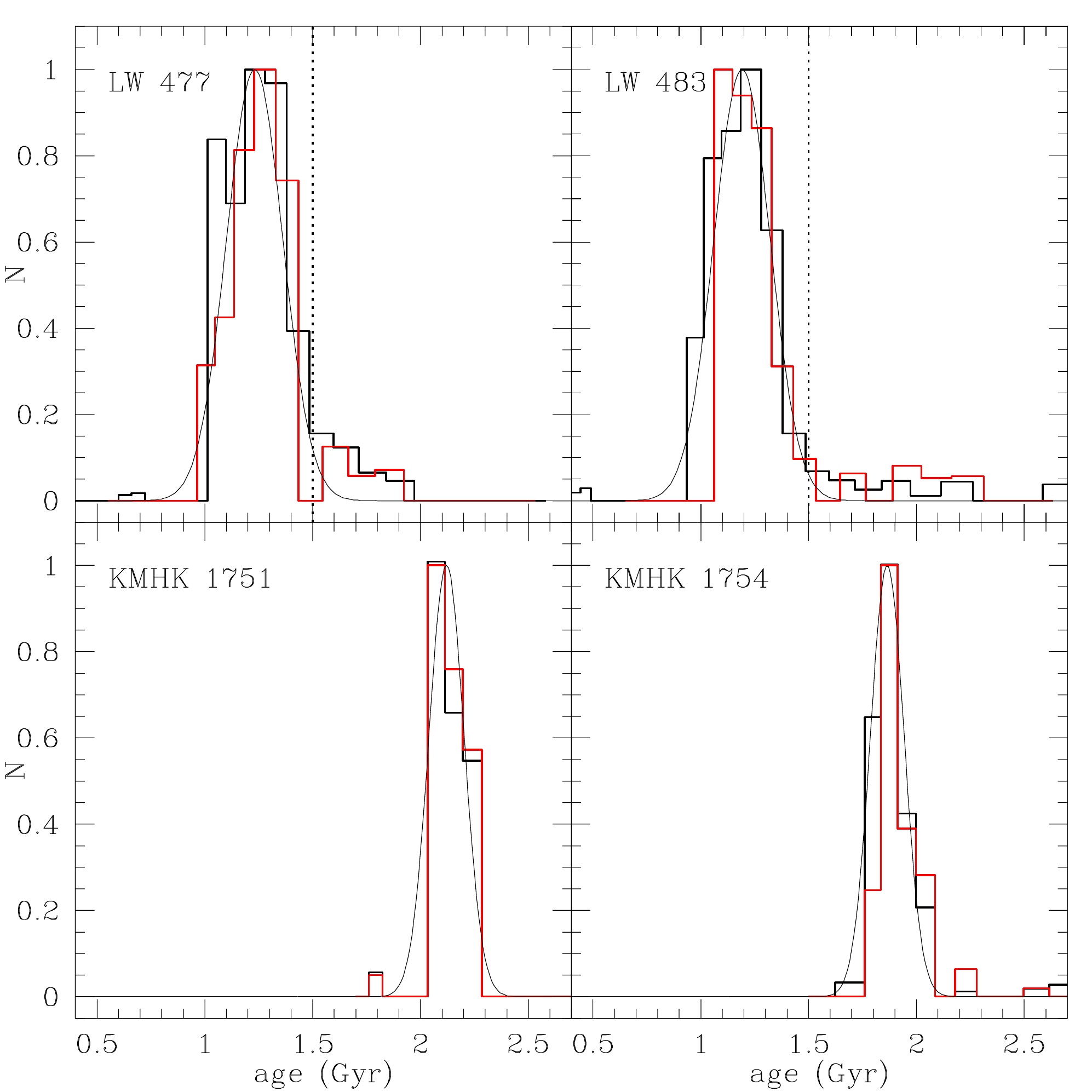}
    \caption{Normalized star counts along the long axis of stars placed within the rectangles drawn 
in Fig.~\ref{fig:fig4}. Thick black and red histograms refer to those obtained by considering or not
the photometric uncertainties, respectively. Gaussian distributions fitted to the black histograms
are also superimposed.}
    \label{fig:fig5}
\end{figure*}

\subsection{Cluster eMSTOs}

\subsubsection{Distribution Across the MSTO}
In order to quantify the MSTO extensions we counted the number of stars (N) located within the parallelograms
drawn in Fig.~\ref{fig:fig4} (the clusters CMDs cleaned from field star contamination) using as independent variable the coordinate (X) along the long axis, according
to the precepts outlined by \citet[][see their figure 14]{goudfrooijetal2011} and used from similar
photometry by \citet{p13}. 
When building the histogram we have taken into account the effects caused by using different binning as well
as the photometric uncertainties, which cause the points have the chance to fall outside their own bin if they 
do not fall in the bin centre \citep[see, e.g.][]{p10,p14b,pg13,petal14c}; thus producing a real distribution. 
For our purposes, we first considered the X range split in bins with sizes 0.10 mag. On the other hand, 
each X point with its error ($\sigma$(X)) 
covers a segment whose size is given by 2$\times$$\sigma$(X), and may or may not fall centred on 
one of the bins, and has dimensions smaller, similar or larger than the bin wherein it is 
placed. These scenarios generate a variety of possibilities, in the sense that the X segment could 
cover from one up to 5 bins depending on its position and size. For this reason, we weighed the 
contribution of each X point to each one of the bins occupied by it, so that the sum of all the 
weights equals unity. The assigned weight was computed as the fraction of its X segment
[2$\times$$\sigma$(X)] that falls in the bin. In practice, we focused on a single
bin and computed the weighted contribution of all the X points to that bin. Then we repeated the
calculation for all the bins. The challenge of knowing whether a portion of an X point (an X 
segment strictly speaking) falls in an bin, was solved by taking into account the following 
possibilities of combination between them.  For each interval we looked for X values
that fall inside the considered bin, as well as X points where X $\pm$ $\sigma$(X)
could cause them to fall in the considered bin.  Note that if X 
$\pm$ $\sigma$(X) causes a X point to step over the considered bin, then we consider that that
X point may have a value that places it inside the considered
interval as well. Therefore, we weighed the contribution of each X point due to their 
point spread functions. We are confident that
our analysis yields accurate morphology and position of the main
features in the resulting X distributions.

The resulting histograms are depicted in Fig.~\ref{fig:fig5} with thick black lines, 
which clearly exhibit extended 
distributions, particularly for the two younger clusters. Note that we have normalized the 
histograms for comparison purposes and avoided the binary sequences, which should be placed
to the right of the vertical dotted lines.
%Nevertheless, the maximum number of stars (N$_{\rm max}$) 
%counted for each cluster is listed in Table~\ref{tab:table6}. 
We then fitted Gaussian functions to 
Fig.~\ref{fig:fig5} using the IRAF {\sc ngaussfit} task.  Note that these histograms account 
for the real spread caused by photometric uncertainties. Whenever the latter are not
taken into account, we obtain very similars distribuions (see histograms drawn with thick red lines),
which suggests that  the observed age spread is not driven by photometric
errors. The estimated age spreads (including the effects of photometric uncertainties), estimated from 
the $FWHM$s are listed in Table~\ref{tab:table6}. 

We note that the two younger clusters (LW~477 and LW~483, especially LW483) have relatively extended MSTOs,
with inferred age spreads of $\sim300$~Myr.  In contrast, the two older cluster (KMHK~1751 and 1754)
have lower inferred age spreads ($<200$~Myr). Note that for a given level of photometric
error and binarity, and assuming SSPs for all the clusters, we would expect the age spread to increase in the older ones, which is not the case for LW~447 and 483.

\subsubsection{Synthetic Cluster Simulations}

In addition to the above tests we also created synthetic clusters, under the assumption that they are pure SSPs, and compared them with the observations.  The goal of this test is to determine whether the clusters do in fact have MSTOs that are more extended than would be extended for an SSP with photometric errors and binaries in the population.

For each cluster we adopted an age (and extinction/distance modulus) from the previous analyses, and constructed an SSP based on model isochrones.  We included errors from the observations and a Salpeter~(1955) IMF above $1$~\msun.  We sampled the IMF stochastically and included photometric uncertainties taken from the observations.  Additionally, we included binaries in our synthetic cluster which we sampled from a flat mass ratio distribution, with a variable binary fraction.

In order to match the synthetic clusters to observations, we took two cuts across the data/models.  First, we 'verticalised' the main sequence, just below the MSTO and made a histogram of the verticalised colour of observed stars and synthetic populations in a 0.5 magnitude bin.  We adjusted the binary fraction until the observations and models had similar widths (approximated by fitting a Gaussian distribution to the histograms).  The number of stars in the synthetic clusters was matched to the observations in this bin.  We note that we did not use stars in the observed clusters beyond the mass-ratio = 1 line, as these stars are likely old field stars that have not been subtracted with the background.  For LW 477 and LW 483 we find binary fractions of 0.6 and 0.35, respectively (within the mass ratio range of $0.5 - 1.0$).

\begin{figure}
	\includegraphics[width=\columnwidth]{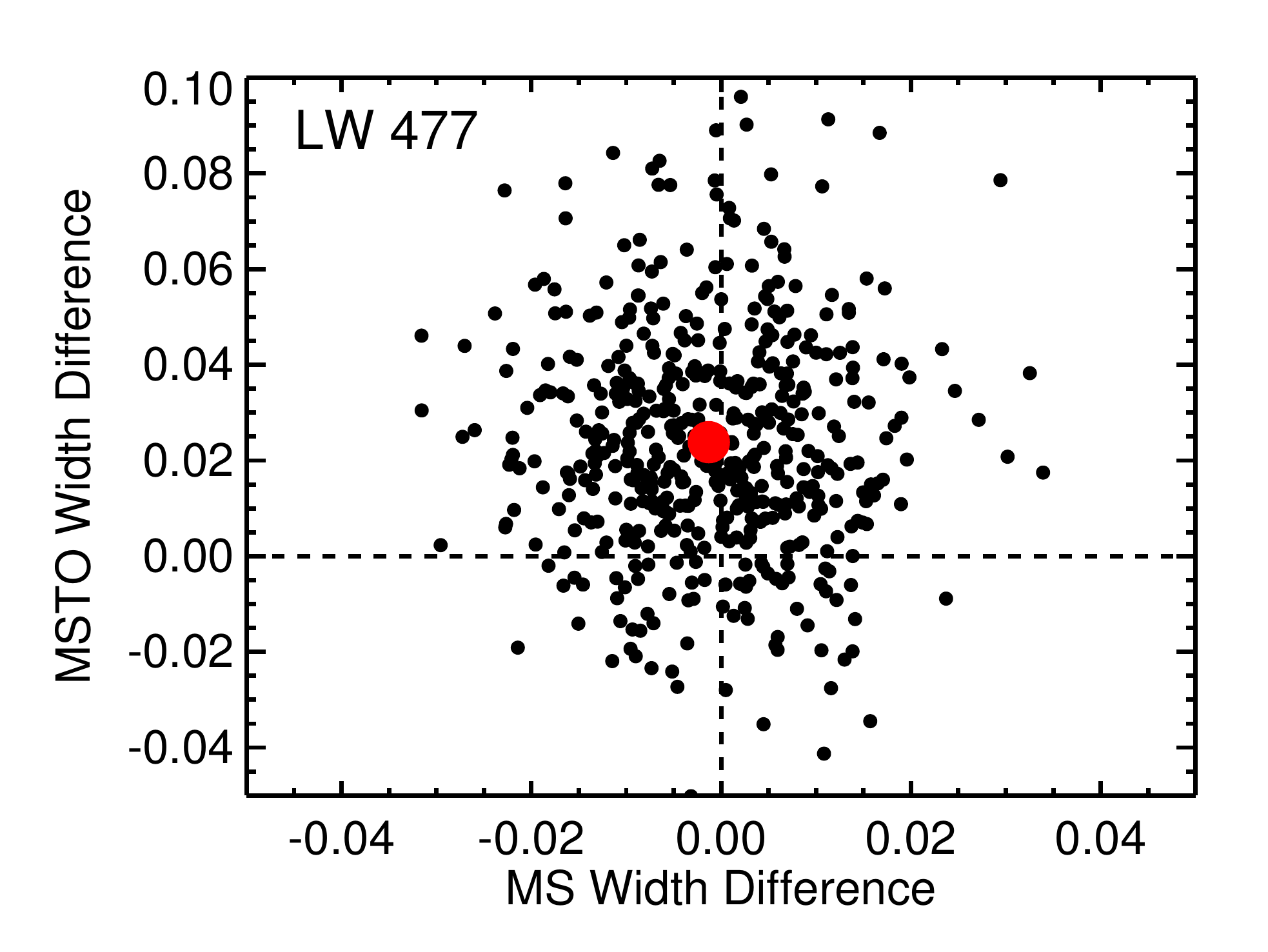}
		\includegraphics[width=\columnwidth]{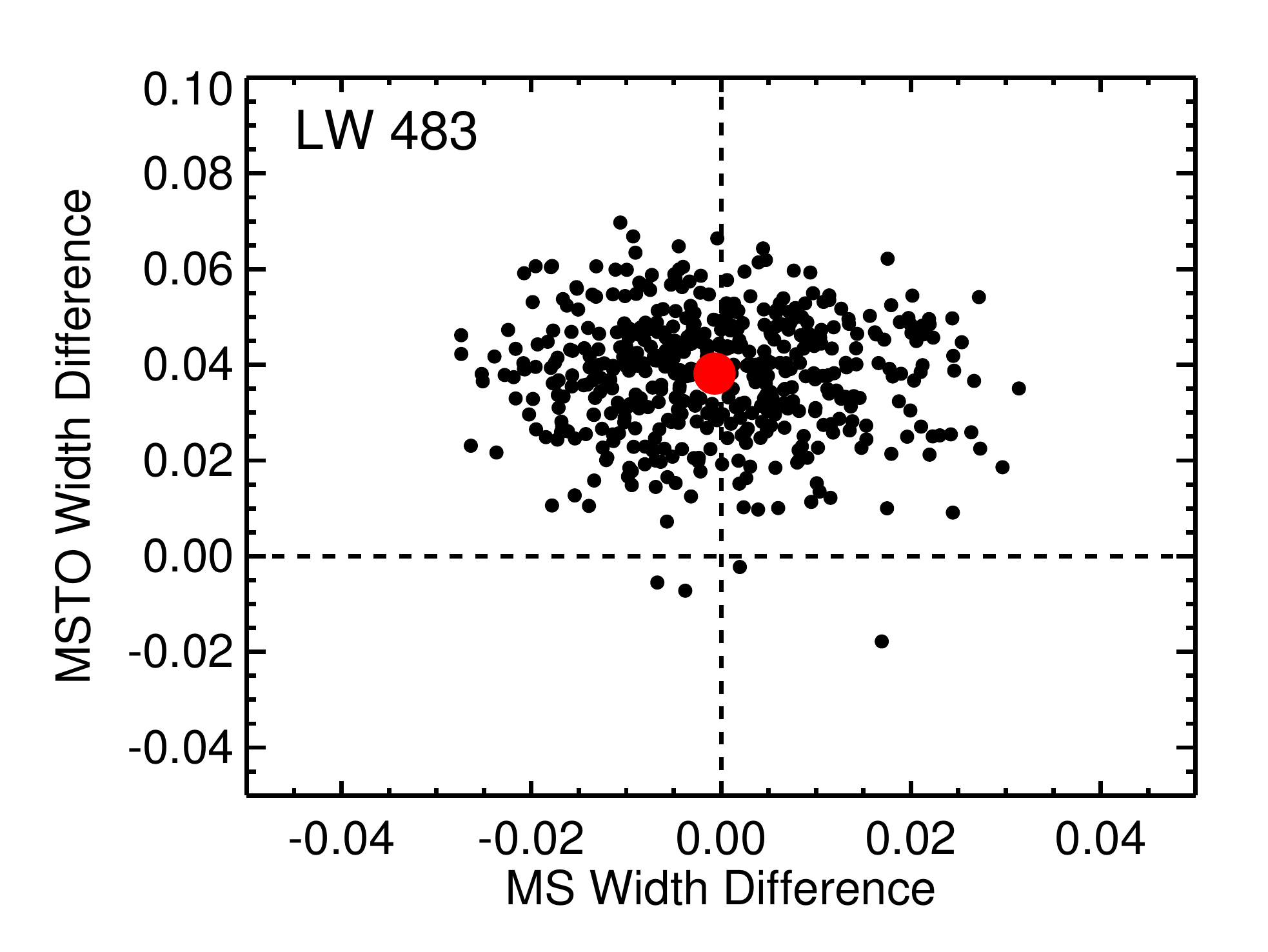}
    \caption{Results of synthetic cluster experiments for LW~477 (top) and LW~483 (bottom).  Once the binary populations of each cluster has been estimate (see text for details) we ran 500 stochastic realisations of synthetic clusters with those properties.  We then took cuts across the (verticalised) main sequence (MS) and main-sequence turn-off (MSTO), fit the resultant histograms with Gaussian functions, and plotted the difference between the observations and synthetic cluster widths for the MS (x-axis) and MSTO (y-axis).  The large (filled) red circle represents the mean of the 500 realisations.  The fact that in both clusters the MS width between the simulations and observations shows that the clusters are well reproduced by the adopted population properties.  However, in both clusters the MSTO is wider in the observations than expected from the simulations, showing that the clusters do indeed host extended MSTOs.}
    \label{fig:fig6}
\end{figure}

Once the binary fractions were fixed, we then took a cut across the MSTO as discussed above and looked at the histogram of observed/synthetic stars across this cut.  Again, the width of the observed distribution and the synthetic populations were approximated by fitting a Gaussian to the distributions. For each cluster we carried out 500 realisations of synthetic clusters with the determined properties (i.e., binary fraction and photometric errors), and compared the widths of the MS (verticalised) and MSTO.  The results for LW 477 and 483 are shown in Fig.~\ref{fig:fig6}.  The x and y-axes of the panels represents the width difference between the observed and synthetic along the MS and MSTO, respectively.

The fact that the width of the MS for both clusters is similar to the width of the synthetic clusters shows that the stellar populations (binaries and photometric errors) are consistent.  On the other hand, the width of the MSTO is larger in both clusters than in average synthetic population.  From this we concluded that both LW~477 and LW~483 host extended MSTOs, which are not caused by binarity or photometric errors.  However, we note that the observed eMSTO in LW~477 may be consistent with no spread, whereas the spread in LW~483 is much more reliably measured.

On the other hand we found that both KMHK~1751 and 1754 have MSTOs consistent with expectations of an SSP (plus binaries and photometric errors).  Hence, the derived extent of the MSTO in the previous section is an upper limit to their true extent.

\section{Cluster structural properties}

We determined the geometrical centres of the clusters in order to obtain their stellar density radial 
profiles.
%circular extracted CMDs where the 
%fiducial features of the clusters could be clearly seen. 
The coordinates of the cluster centres and their estimated uncertainties were determined by fitting Gaussian 
distributions to the star counts in the $x$ and $y$ directions for each cluster. The fits of the Gaussians
 were 
performed using the {\sc ngaussfit} routine in the {\sc stsdas/iraf} package. We adopted a single Gaussian
 and fixed the constant 
to the corresponding background levels (i.e. stellar field densities assumed to be uniform) and the linear 
terms to zero. The centre of the Gaussian, its amplitude and its $FWHM$ acted as variables. 
The number of stars projected along the $x$ and $y$ directions were counted within intervals of 40 pixel wide. 
In addition, we checked that using spatial bins from 20 to 60 pixels does not result in 
significant changes in the derived centres. 
Cluster centres were finally determined with a typical standard deviation of 
$\pm$ 10 pixels  ($\sim$ 0.4 pc) in all cases. Likewise, we took advantage of the deepest $g,i$ images
obtained for each cluster to estimate their centres using the {\sc stsdas.n2gaussfit} routine.
During the fitting procedure, we allowed for the ellipticity, the position angle, the peak amplitude,
and the $FWHM$ to vary from their initial guesses, and kept fixed the background level. From both $g,i$
images we recovered  values closer than $\pm$10 pixels from the cluster centres estimated
from star counts, which implies that there is no bright star affecting their innermost light profiles. 

We then used the resulting centres to checked whether the clusters exhibit some elliptical signature,
even though we obtained from {\sc n2gaussfit} ellipticities equals to zero.
To do that, we employed the {\sc stsdas.ellipse} task, allowing only for the ellipticity and the 
position angle to change (i.e., we held cluster centres fixed). We tried different combinations of
first guesses for the ellipticity and the position angle, as well as different steps between
successive ellipses, initial semi-major lenght, etc. in order to be certain that the convergency
of the elliptical isophotes fitted at different distances to the clusters' centres did not depend on 
those values. The large number of configurations performed led us to conclude that the four clusters
do not have any measurable sign of ellipticity.

\begin{figure*}
	\includegraphics[width=\textwidth]{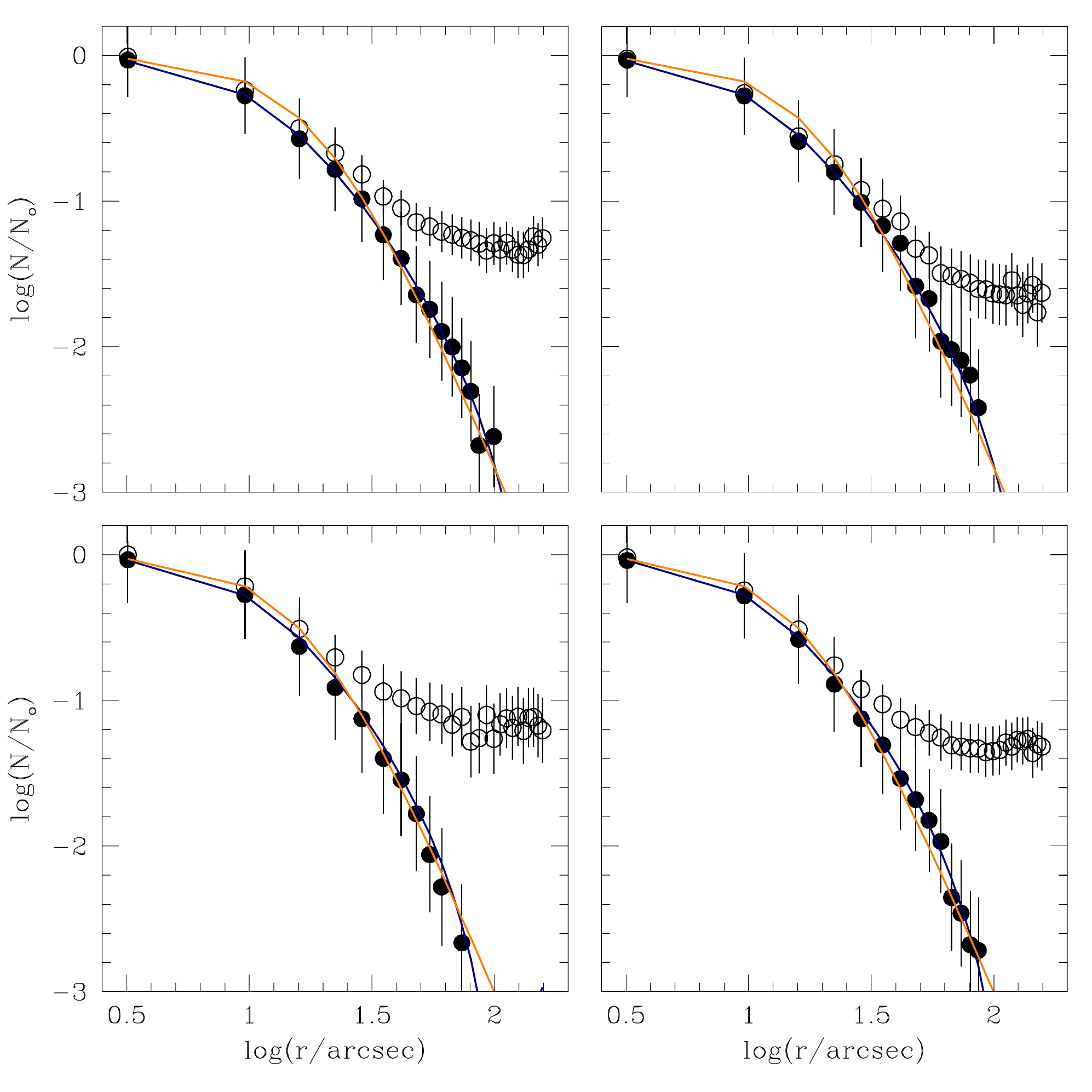}
    \caption{Stellar density cluster profiles obtained from star counts for LW\,477 (upper-left),
LW\,483 (upper-right), KMHK\,1751 (bottom-left) and KMHK\,1754 (bottom-right).Open and filled circles 
refer to measured and
background subtracted surface brightness profiles, respectively. Blue and orange solid lines
depict the fitted King and Plummer curves, respectively.}
    \label{fig:fig7}
\end{figure*}

We also built stellar density profiles based on completeness corrected star counts previously performed within boxes of 40 pixels a side distributed 
throughout the whole field of each cluster. The selected size of the box allowed us to sample statistically 
the stellar spatial distribution. 
Thus, the number of stars per unit area at a given radius, $r$, can be directly calculated through 
the expression:

\begin{equation}
(n_{r+20} - n_{r-20})/(m_{r+20} - m_{r-20}),
\end{equation}

\noindent where $n_j$ and $m_j$ represent the number of stars and boxes included in a circle of radius $j$, 
respectively. Note that this method does not necessarily require a complete circle of radius $r$ within the
 observed 
field to estimate the mean stellar density at that distance. This is an important consideration since having a stellar 
density profile which extends far away from the cluster centre allows us to estimate the background level with 
high precision. This is necessary to derive the cluster radius ($r_{cls}$). The resulting density profiles 
expressed as number of stars per arcsec$^2$ are shown in  Fig.~\ref{fig:fig7}.
In the figure, we 
%indicated the background level with a dotted line and 
represented the constructed and
background subtracted stellar density profiles with open and filled circles, respectively.
Errorbars represent Poisson errors, to which we added the rms error of the
background star count to the background subtracted density profiles.

The background corrected surface density profiles were fitted using a \citet{king62}'s model through the expression :

\begin{equation}
 N \varpropto ({\frac{1}{\sqrt{1+(r/r_c)^2}} - \frac{1}{\sqrt{1 + (r_t/r_c)^2}}})^2
\end{equation}

\noindent where 
%$A$ is a constant, and 
$r_c$ and $r_t$ are the core and tidal radii, respectively. The values derived for $r_c$ and 
$r_t$ from the fits are listed in Table~\ref{tab:table7}, while the respective King's curves are plotted with blue solid lines in Fig.~\ref{fig:fig7}.
%and \ref{fig:fig8}. 
As can be seen, the King profiles satisfactorily reproduce the whole cluster extensions.
Nevertheless, in order to get independent estimates of the cluster half-mass radii, we
%regions and slightly underestimate the cluster extensions, a behaviour found in clusters not tidally truncated
%\citep{miocchietal13,kimetal15,dalessandroetal15}. In order to account for such extra-tidal stars we 
fitted Plummer's profiles using the expression:

\begin{equation}
N \varpropto \frac{1}{(1+(r/a)^2)^2} 
\end{equation}

\noindent where $a$ is the Plummer's radius, which is related to the half-mass radius ($r_h$) by the relation $r_h$ $\sim$ 1.3$a$. The resulting $r_h$ values are listed in Table~\ref{tab:table7} and the corresponding Plummer's curves drawn with orange solid lines in Fig.~\ref{fig:fig7}.

% and \ref{fig:fig8}. These curves slightly improve the fitting of the
%outer cluster regions.

%The mean solutions
%found of the King and Plummer {\bf models
% from the fit of the clusters surface brightness profiles 
%were} superimposed using blue and orange solid lines, respectively. As can be seen, there is an overall good agreement.

%\begin{table}
%\caption{King and Plummer parameters of LMC star clusters.}
%\label{tab:table7}
%\begin{tabular}{@{}lcccc}\hline
%Star cluster & image & $r_c$  & $r_t$  & $r_h$\\
%             &    & (arcsec) & (arcsec) & (arcsec)\\\hline
%LW\,477      &  $g$    & 4.5$\pm$0.5 & 45$\pm$5 & 13.0$\pm$0.5 \\ 
%             &   $i$    & 4.5$\pm$0.5 & 45$\pm$5 & 13.0$\pm$0.5\\
%LW\,483      &     $g$  & 4.5$\pm$0.5 & 45$\pm$5 & 13.0$\pm$0.5\\
%             &    $i$   & 4.5$\pm$0.5 & 45$\pm$5 & 13.0$\pm$0.5\\
%KMHK1751     &   $g$  & 5.5$\pm$0.5 & 20$\pm$5 & 8.5$\pm$0.5\\
%             &    $i$    & 5.0$\pm$0.5 & 10$\pm$5 & 8.0$\pm$0.5 \\
%KMHK1754     & $g$   & 6.0$\pm$0.5 & 15$\pm$5 & 9.8$\pm$0.5\\
%             &    $i$    & 6.0$\pm$0.5 & 15$\pm$5 & 9.8$\pm$0.5 \\
%\hline
%\end{tabular}

%\noindent Note: to convert 1 arcsec to pc, use the following expression :
%10$\times$10$^{(m-M)_o/5}$sin(1/3600), where $(m-M)_o$ is the LMC distance modulus.
%\end{table}

\begin{table*}
\centering
\caption{Structural parameters of LMC star clusters.}
\label{tab:table7}
\begin{tabular}{@{}lcccccc}\hline\hline
Star cluster & $d_{GC}$ &  $r_c$ & $r_h$ &  $r_{cls}$ &  $r_t$ & $r_J$  \\
             &  (kpc)   &   (pc) &  (pc) &   (pc)     &  (pc)  & (pc) \\\hline
LW\,477      &   6.1    & 2.7$\pm$0.5 & 6.3$\pm$0.6 & 18.6$\pm$0.9 & 36.4$\pm$2.4 & 21.6\\
LW\,483      &   6.7    & 2.7$\pm$0.5 & 6.3$\pm$0.6 & 21.8$\pm$1.2 & 36.4$\pm$2.4 & 32.0\\
KMHK1751     &   6.4    & 2.7$\pm$0.5 & 5.7$\pm$0.6 & 15.0$\pm$0.7 & 26.7$\pm$2.4 & 18.4\\
KMHK1754     &   6.1    & 2.7$\pm$0.5 & 5.7$\pm$0.6 & 15.3$\pm$0.9 & 29.1$\pm$2.4 & 18.3\\
\hline
\end{tabular}

\noindent Note: to convert 1 arcsec to pc, we use the following expression :\\
10$\times$10$^{(m-M)_o/5}$sin(1/3600), 
where $(m-M)_o$ = 18.5 mag is the LMC distance modulus.

\end{table*}

The masses of the clusters in our sample were derived by comparing the observed integrated magnitude 
of each cluster (corrected for distance modulus and extinction) to that of the \cite{metal08} SSP 
models with a \cite{kroupa02} stellar initial mass function and a metallicity of $Z=0.008$ of the 
appropriate age.  The errors in the mass estimate are driven largely by the choice of models (which were 
also used in estimating the ages).  We estimate the uncertainty in the mass to be 
$\sigma(\log(M_{cls}/\msun))$ $\sim0.2$~dex, i.e., about a factor of $\sim1.5$.

From the derived masses we estimated both the Jacobi tidal radius and the half-mass relaxion time of 
each cluster. The former was computed from the expression \citep{cw90}:

\begin{equation}
r_J = (\frac{M_{cls}}{3 M_{gal}})^{1/3}\times d_{GC}
\end{equation}

\noindent where $M_{cls}$ is the total cluster mass (see Table~\ref{tab:table6}), $M_{gal}$ is the LMC mass 
inside 8.7 kpc ((1.7$\pm$0.7)$\times$10$^{10}$ M$_{\odot}$,\citet{vdmk14}), and $d_{GC}$ is the cluster
deprojected galactocentric distance. We computed $d_{GC}$ by assuming that the clusters are part of 
a disc having an inclination $i$ = 35.8$\degr$ and a position angle of the line 
of nodes of $\Theta$ = 145$\degr$ \citep{os02}. We refer the reader to Table 1 of \citet{ss10} which includes 
a summary of orientation measurements of the LMC disc plane, as well as their analysis of the orientation and
 other LMC disc quantities, supporting the present adopted values. Table~\ref{tab:table7} lists the
resulting $d_{GC}$ and $r_J$ values. By comparing the $r_J$ values with the cluster tidal radii (see Table 7)
we conclude that the four clusters are not tidally truncated, i.e., they are not limited.
This means that the clusters are not expected to have lost significant amounts of stellar mass, so that they 
current masses should reflect their initial masses.

On the other hand, for the half-mass relaxation times we used the equation \citep{sh71}:

\begin{equation}
t_r = \frac{8.9\times 10^5 M_{cls}^{1/2} r_h^{3/2}}{\bar{m} log_{10}(0.4M_{cls}/\bar{m})}
\end{equation}

\noindent where $M_{cls}$ is the cluster mass (see Table~\ref{tab:table6}), $r_h$ is the half-mass
radius (see Table~\ref{tab:table7}) and $\bar{m}$ is the average mass of the cluster stars.  For simplicity we assume a constant 
average stellar mass of 1 M$_{\odot}$.
%Here, we estimated the average stellar mass as $\bar{m}$ $\sim$ 1 M$_{\odot}$. 
The derived relaxation times are
listed in  Table~\ref{tab:table6}. These values are in excellent agreement with those coming from
using Fig. 20 of \citet{pijlooetal15}. Indeed, we found no difference between them ($\Delta$($t_r$) =
1.3$\pm$9.5 Myr), independently if the latter were interpolated by considering scenarios with and without 
mass segregation. Additionally, we found a half-mass density range of 6-16 M$_{\odot}$ pc$^3$ for the
studied clusters. These value are much larger than the minimum density a cluster needs to have in order 
to be stable against the tidal disruption of a galaxy ($\sim$ 0.1M$_{\odot}$ pc$^{3}$, \citet{bok34}).
Accordingly, \citet{wilkinsonetal03} also showed that the tidal field of the LMC does not cause any 
perturbation on the clusters.

\section{Discussion}

Two low mass clusters studied in the present work show evidence of eMSTOs.  
This feature has now been commonly observed in intermediate age ($1-2$~Gyr) clusters as well as 
younger clusters (NGC~1856, $\sim300$~Myr), although most of the clusters studied to date have 
much higher stellar masses than those in the current study.  The four clusters have masses 
$\lesssim5\times10^3$~\msun, and due to the fact that they are tidally under-filling, are not
 expected to have lost significant amounts of stars due to tidal interactions or 2-body relaxation.

Our results suggest that cluster mass (and correspondingly, cluster escape velocity) is not
 directly linked to the eMSTO phenomena.  This is in contradiction to the scenario of 
\citet{goudfrooijetal14}, who suggested that only clusters with escape velocities $>15$~km/s when
 they were young ($\sim10$~Myr) would be able to retain stellar ejecta and accrete material 
from the cluster surroundings to undergo extended star-formation episodes (lasting hundreds of Myr).
  In this scenario, the inferred age spreads from the eMSTO would correspond to actual age spreads
 within the clusters.

On the other hand, our results confirm recent suggestions that the inferred age spreads found in 
clusters is related to the age of the cluster, with younger clusters having small inferred age
 spreads, and older clusters showing larger age spreads, with a peak at $\sim1.5$~Gyr and the 
inferred age spread decreasing after this until it disappears around $\sim2$~Gyr 
\citep{bh15b,niederhoferetal15b,niederhoferetal15c}.  This strong correlation with cluster age 
suggests that a stellar evolutionary effect is the underlying cause, and one potential mechanism 
is stellar rotation \citep{bdm09,bh15b,niederhoferetal15b}.

\section{Conclusions}

We have analysed deep $g,i$ images of four low mass clusters in the outer regions of the LMC.  
We have used the images to derive the structural parameters of the clusters. We built
their number density radial profiles from star counts carried out throughout the observed
fields using the final photometric catalogues. Then, we fitted King and Plummer models to derive
cluster core, half-mass and tidal radii. 
%The solutions obtained tightly reproduce the cluster stellar
%radial profiles built from star counts using the final photometric catalogues. 

The four clusters were found to be of relatively small size ($<$ 22 pc) and are tidally under-filling  as well;
their Jacobi radii being larger than the cluster radii. The four clusters have core radii of  2.7 pc, which are smaller than the values expected for 
intermediate age massive clusters with eMSTOs \citep[e.g.][]{mackeyetal08b,kelleretal11}. Indeed,
\cite{kelleretal12} proposed that the eMSTOs are a common pathway for massive stellar clusters. 
The studied cluster sample seem to be dynamically evolved, since their
estimated relaxation times are many times smaller than the cluster ages (age/$t_r$ $\sim$ 4-13). 
%This means 
%that no further two-body relaxation processes should take place within the clusters. 

One of the goals of the present work was to search for the eMSTOs, which
 are now a commonly observed feature of young ($100-300$~Myr) and intermediate age ($1-2$~Gyr) massive
 clusters in the LMC/SMC.  We have constructed $g-i$ versus $g$ CMDs 
and fit isochrones to the data after accounting for the distance modulus and extinction.  We find that 
LW\,477 and LW\,483 have similar ages, $1.25$~Gyr, while clusters KMHK\,1751 and 1754 are older, both with 
an age of $\sim2$~Gyr.

We find that the two young clusters (LW\,477 and 483) show  eMSTOs with inferred age spreads 
(corrected for photometric errors) of $\sim300$~Myr.  The eMSTO in LW 483 appears to be more robust than LW 477.  These inferred age spreads are  in 
 agreement with those of much more massive clusters of similar age 
\citep[e.g.][ $\sim$ 400-500 Myr]{rubeleetal2010,rubeleetal2011,goudfrooijetal2011,niederhoferetal15c}. 
 This is the first time that eMSTOs have been observed in such low mass clusters, the discovery of which, 
suggests that mass is not an important factor in controlling whether clusters host eMSTOs.  This is in 
contradiction to the scenario of \cite{goudfrooijetal14} who suggest mass (in fact, escape velocity) is 
the controlling factor in the eMSTO phenomenon.

For the two older clusters in our sample (KMHK\,1751 and 1754), we find that both are consistent with a
 small age spread ($<200$~Myr), much smaller than that found for the two young clusters. This is
 consistent with the stellar rotation scenario, which predicts that after $1-1.5$~Gyr the inferred age 
spreads should decrease as lower mass stars (which are affected by magnetic breaking) enter the MSTO
\citep{bh15b,niederhoferetal15b}.

Our observations have explicitly tested the role of cluster mass in the eMSTO phenomenon, and we find 
that it does not play a significant part.  Taken together with other recent results in the literature
\citep{lietal14,bh15a,bh15b,niederhoferetal15,niederhoferetal15b,niederhoferetal15c}, our results
 disfavour 
actual age spreads as the origin of the eMSTO phenomenon, and favour a stellar evolutionary effect,
 such as
 stellar rotation. While stellar rotation may explain the MSTO phenomenon, high-precision CMDs of younger clusters have revealed features (e.g., dual main sequences) that may be difficult to explain within the rotational scenario.
The present results point to the need of high precision HST imaging 
of these and other low mass intermediate age clusters to further constrain the populations within 
the clusters.

\begin{acknowledgements}
NB gratefully acknowledges financial support from the Royal Society (University Research Fellowship) and 
the European Research Council (ERC-CoG-646928, Multi-Pop).  Based on observations obtained at the Gemini 
Observatory (Programme: GS-2015A-Q-44), which is operated by the
Association of Universities for Research in Astronomy, Inc., under a cooperative agreement
with the NSF on behalf of the Gemini partnership: the National Science Foundation (United
States), the Science and Technology Facilities Council (United Kingdom), the National Research Council (Canada), 
CONICYT (Chile), the Australian Research Council (Australia),
Minist\'erio da Ci\`encia, Tecnologia e Inova\c{c}\~ao (Brazil) and Ministerio de Ciencia, Tecnolog\'{\i}a e
Innovaci\'on Productiva (Argentina).  
 We thank the anonymous referee whose thorough comments and suggestions
allowed us to improve the manuscript.
\end{acknowledgements}

\bibliographystyle{aa}
%\bibliography{paper} % if your bibtex file is called paper.bib

\begin{thebibliography}{59}
\expandafter\ifx\csname natexlab\endcsname\relax\def\natexlab#1{#1}\fi

\bibitem[{{Bastian} {et~al.}(2013){Bastian}, {Cabrera-Ziri}, {Davies}, \&
  {Larsen}}]{bastianetal13}
{Bastian}, N., {Cabrera-Ziri}, I., {Davies}, B., \& {Larsen}, S.~S. 2013,
  \mnras, 436, 2852

\bibitem[{{Bastian} \& {de Mink}(2009)}]{bdm09}
{Bastian}, N. \& {de Mink}, S.~E. 2009, \mnras, 398, L11

\bibitem[{{Bastian} \& {Silva-Villa}(2013)}]{bsv13}
{Bastian}, N. \& {Silva-Villa}, E. 2013, \mnras, 431, L122


\bibitem[{{Bastian} {et~al.}(2016)}]{bastianetal2016}
{Bastian}, N. {et~al.} 2016, \mnras, submitted

\bibitem[{{Baumgardt} {et~al.}(2013){Baumgardt}, {Parmentier}, {Anders}, \&
  {Grebel}}]{baetal13}
{Baumgardt}, H., {Parmentier}, G., {Anders}, P., \& {Grebel}, E.~K. 2013,
  \mnras, 430, 676

\bibitem[{{Bertelli} {et~al.}(2003){Bertelli}, {Nasi}, {Girardi}, {Chiosi},
  {Zoccali}, \& {Gallart}}]{bertellietal2003}
{Bertelli}, G., {Nasi}, E., {Girardi}, L., {et~al.} 2003, \aj, 125, 770

\bibitem[{{Bica} {et~al.}(2008){Bica}, {Bonatto}, {Dutra}, \&
  {Santos}}]{betal08}
{Bica}, E., {Bonatto}, C., {Dutra}, C.~M., \& {Santos}, J.~F.~C. 2008, \mnras,
  389, 678

\bibitem[{{Bok}(1934)}]{bok34}
{Bok}, B.~J. 1934, Harvard College Observatory Circular, 384, 1

\bibitem[{{Brandt} \& {Huang}(2015{\natexlab{a}})}]{bh15b}
{Brandt}, T.~D. \& {Huang}, C.~X. 2015{\natexlab{a}}, \apj, 807, 25

\bibitem[{{Brandt} \& {Huang}(2015{\natexlab{b}})}]{bh15a}
{Brandt}, T.~D. \& {Huang}, C.~X. 2015{\natexlab{b}}, \apj, 807, 24

\bibitem[{{Cabrera-Ziri} {et~al.}(2014){Cabrera-Ziri}, {Bastian}, {Davies},
  {Magris}, {Bruzual}, \& {Schweizer}}]{cabrerazirietal14}
{Cabrera-Ziri}, I., {Bastian}, N., {Davies}, B., {et~al.} 2014, \mnras, 441,
  2754

\bibitem[{{Cabrera-Ziri} {et~al.}(2015){Cabrera-Ziri}, {Bastian}, {Longmore},
  {Brogan}, {Hollyhead}, {Larsen}, {Whitmore}, {Johnson}, {Chandar}, {Henshaw},
  {Davies}, \& {Hibbard}}]{cabrerazirietal15}
{Cabrera-Ziri}, I., {Bastian}, N., {Longmore}, S.~N., {et~al.} 2015, \mnras,
  448, 2224

\bibitem[{{Cardelli} {et~al.}(1989){Cardelli}, {Clayton}, \&
  {Mathis}}]{cetal89}
{Cardelli}, J.~A., {Clayton}, G.~C., \& {Mathis}, J.~S. 1989, \apj, 345, 245

\bibitem[{{Chernoff} \& {Weinberg}(1990)}]{cw90}
{Chernoff}, D.~F. \& {Weinberg}, M.~D. 1990, \apj, 351, 121

\bibitem[{{Conroy} \& {Spergel}(2011)}]{cs11}
{Conroy}, C. \& {Spergel}, D.~N. 2011, \apj, 726, 36

\bibitem[{{Correnti} {et~al.}(2015){Correnti}, {Goudfrooij}, {Puzia}, \& {de
  Mink}}]{correntietal15}
{Correnti}, M., {Goudfrooij}, P., {Puzia}, T.~H., \& {de Mink}, S.~E. 2015,
  \mnras, 450, 3054

\bibitem[{{D'Antona} {et~al.}(2015){D'Antona}, {Di Criscienzo}, {Decressin},
  {Milone}, {Vesperini}, \& {Ventura}}]{dantonaetal15}
{D'Antona}, F., {Di Criscienzo}, M., {Decressin}, T., {et~al.} 2015, \mnras,
  453, 2637

\bibitem[{{de Grijs} {et~al.}(2014){de Grijs}, {Wicker}, \& {Bono}}]{dgetal14}
{de Grijs}, R., {Wicker}, J.~E., \& {Bono}, G. 2014, \aj, 147, 122

\bibitem[{{Dutra} {et~al.}(2001){Dutra}, {Bica}, {Clari{\'a}}, {Piatti}, \&
  {Ahumada}}]{dutraetal01}
{Dutra}, C.~M., {Bica}, E., {Clari{\'a}}, J.~J., {Piatti}, A.~E., \& {Ahumada},
  A.~V. 2001, \aap, 371, 895

\bibitem[{{Ekstr{\"o}m} {et~al.}(2012){Ekstr{\"o}m}, {Georgy}, {Eggenberger},
  {Meynet}, {Mowlavi}, {Wyttenbach}, {Granada}, {Decressin}, {Hirschi},
  {Frischknecht}, {Charbonnel}, \& {Maeder}}]{ekstrometal12}
{Ekstr{\"o}m}, S., {Georgy}, C., {Eggenberger}, P., {et~al.} 2012, \aap, 537,
  A146

\bibitem[{{Gao} {et~al.}(2013){Gao}, {Jiang}, {Li}, \& {Xue}}]{getal13}
{Gao}, J., {Jiang}, B.~W., {Li}, A., \& {Xue}, M.~Y. 2013, \apj, 776, 7

\bibitem[{{Georgy} {et~al.}(2014){Georgy}, {Granada}, {Ekstr{\"o}m}, {Meynet},
  {Anderson}, {Wyttenbach}, {Eggenberger}, \& {Maeder}}]{georgyetal14}
{Georgy}, C., {Granada}, A., {Ekstr{\"o}m}, S., {et~al.} 2014, \aap, 566, A21

\bibitem[{{Glatt} {et~al.}(2009){Glatt}, {Grebel}, {Gallagher}, {Nota},
  {Sabbi}, {Sirianni}, {Clementini}, {Da Costa}, {Tosi}, {Harbeck}, {Koch}, \&
  {Kayser}}]{getal09}
{Glatt}, K., {Grebel}, E.~K., {Gallagher}, III, J.~S., {et~al.} 2009, \aj, 138,
  1403

\bibitem[{{Goudfrooij} {et~al.}(2014){Goudfrooij}, {Girardi},
  {Kozhurina-Platais}, {Kalirai}, {Platais}, {Puzia}, {Correnti}, {Bressan},
  {Chandar}, {Kerber}, {Marigo}, \& {Rubele}}]{goudfrooijetal14}
{Goudfrooij}, P., {Girardi}, L., {Kozhurina-Platais}, V., {et~al.} 2014, \apj,
  797, 35

\bibitem[{{Goudfrooij} {et~al.}(2011){Goudfrooij}, {Puzia},
  {Kozhurina-Platais}, \& {Chandar}}]{goudfrooijetal2011}
{Goudfrooij}, P., {Puzia}, T.~H., {Kozhurina-Platais}, V., \& {Chandar}, R.
  2011, \apj, 737, 3

\bibitem[{{Keller} {et~al.}(2011){Keller}, {Mackey}, \& {Da
  Costa}}]{kelleretal11}
{Keller}, S.~C., {Mackey}, A.~D., \& {Da Costa}, G.~S. 2011, \apj, 731, 22

\bibitem[{{Keller} {et~al.}(2012){Keller}, {Mackey}, \& {Da
  Costa}}]{kelleretal12}
{Keller}, S.~C., {Mackey}, A.~D., \& {Da Costa}, G.~S. 2012, \apjl, 761, L5

\bibitem[{{King}(1962)}]{king62}
{King}, I. 1962, \aj, 67, 471

\bibitem[{Kroupa(2002)}]{kroupa02}
Kroupa, P. 2002, Science, 295, 82

\bibitem[{{Li} {et~al.}(2014){Li}, {de Grijs}, \& {Deng}}]{lietal14}
{Li}, C., {de Grijs}, R., \& {Deng}, L. 2014, \nat, 516, 367

\bibitem[{{Mackey} \& {Broby Nielsen}(2007)}]{mb07}
{Mackey}, A.~D. \& {Broby Nielsen}, P. 2007, \mnras, 379, 151

\bibitem[{{Mackey} {et~al.}(2008{\natexlab{a}}){Mackey}, {Broby Nielsen},
  {Ferguson}, \& {Richardson}}]{maetal08}
{Mackey}, A.~D., {Broby Nielsen}, P., {Ferguson}, A.~M.~N., \& {Richardson},
  J.~C. 2008{\natexlab{a}}, \apjl, 681, L17

\bibitem[{{Mackey} {et~al.}(2008{\natexlab{b}}){Mackey}, {Wilkinson}, {Davies},
  \& {Gilmore}}]{mackeyetal08b}
{Mackey}, A.~D., {Wilkinson}, M.~I., {Davies}, M.~B., \& {Gilmore}, G.~F.
  2008{\natexlab{b}}, \mnras, 386, 65

\bibitem[{{Magrini} {et~al.}(2009){Magrini}, {Sestito}, {Randich}, \&
  {Galli}}]{metal09}
{Magrini}, L., {Sestito}, P., {Randich}, S., \& {Galli}, D. 2009, \aap, 494, 95

\bibitem[{{Marigo} {et~al.}(2008){Marigo}, {Girardi}, {Bressan}, {Groenewegen},
  {Silva}, \& {Granato}}]{metal08}
{Marigo}, P., {Girardi}, L., {Bressan}, A., {et~al.} 2008, \aap, 482, 883

\bibitem[{{Milone} {et~al.}(2009){Milone}, {Bedin}, {Piotto}, \&
  {Anderson}}]{mietal09}
{Milone}, A.~P., {Bedin}, L.~R., {Piotto}, G., \& {Anderson}, J. 2009, \aap,
  497, 755

\bibitem[{{Milone} {et~al.}(2015){Milone}, {Bedin}, {Piotto}, {Marino},
  {Cassisi}, {Bellini}, {Jerjen}, {Pietrinferni}, {Aparicio}, \&
  {Rich}}]{miloneetal15}
{Milone}, A.~P., {Bedin}, L.~R., {Piotto}, G., {et~al.} 2015, \mnras, 450, 3750

\bibitem[{{Milone} {et~al.}(2016){Milone}, {Marino}, {D'Antona}, {Bedin}, {Da
  Costa}, {Jerjen}, \& {Mackey}}]{miloneetal2016}
{Milone}, A.~P., {Marino}, A.~F., {D'Antona}, F., {et~al.} 2016, ArXiv e-prints
  [\eprint[arXiv]{1603.03493}]

\bibitem[{{Niederhofer} {et~al.}(2015{\natexlab{a}}){Niederhofer}, {Bastian},
  {Kozhurina-Platais}, {Hilker}, {de Mink}, {Cabrera-Ziri}, {Li}, \&
  {Ercolano}}]{niederhoferetal15c}
{Niederhofer}, F., {Bastian}, N., {Kozhurina-Platais}, V., {et~al.}
  2015{\natexlab{a}}, ArXiv e-prints [\eprint[arXiv]{1510.08476}]

\bibitem[{{Niederhofer} {et~al.}(2015{\natexlab{b}}){Niederhofer}, {Georgy},
  {Bastian}, \& {Ekstr{\"o}m}}]{niederhoferetal15b}
{Niederhofer}, F., {Georgy}, C., {Bastian}, N., \& {Ekstr{\"o}m}, S.
  2015{\natexlab{b}}, \mnras, 453, 2070

\bibitem[{{Niederhofer} {et~al.}(2015{\natexlab{c}}){Niederhofer}, {Hilker},
  {Bastian}, \& {Silva-Villa}}]{niederhoferetal15}
{Niederhofer}, F., {Hilker}, M., {Bastian}, N., \& {Silva-Villa}, E.
  2015{\natexlab{c}}, \aap, 575, A62

\bibitem[{{Olsen} \& {Salyk}(2002)}]{os02}
{Olsen}, K.~A.~G. \& {Salyk}, C. 2002, \aj, 124, 2045

\bibitem[{{Piatti}(2010)}]{p10}
{Piatti}, A.~E. 2010, \aap, 513, L13

\bibitem[{{Piatti}(2011)}]{p11a}
{Piatti}, A.~E. 2011, \mnras, 418, L40

\bibitem[{{Piatti}(2013)}]{p13}
{Piatti}, A.~E. 2013, \mnras, 430, 2358

\bibitem[{{Piatti}(2014)}]{p14b}
{Piatti}, A.~E. 2014, \mnras, 437, 1646

\bibitem[{{Piatti} \& {Bica}(2012)}]{pb12}
{Piatti}, A.~E. \& {Bica}, E. 2012, \mnras, 425, 3085

\bibitem[{{Piatti} {et~al.}(2014{\natexlab{a}}){Piatti}, {del Pino},
  {Aparicio}, \& {Hidalgo}}]{petal14c}
{Piatti}, A.~E., {del Pino}, A., {Aparicio}, A., \& {Hidalgo}, S.~L.
  2014{\natexlab{a}}, \mnras, 443, 1748

\bibitem[{{Piatti} \& {Geisler}(2013)}]{pg13}
{Piatti}, A.~E. \& {Geisler}, D. 2013, \aj, 145, 17

\bibitem[{{Piatti} {et~al.}(2014{\natexlab{b}}){Piatti}, {Keller}, {Mackey}, \&
  {Da Costa}}]{petal14}
{Piatti}, A.~E., {Keller}, S.~C., {Mackey}, A.~D., \& {Da Costa}, G.~S.
  2014{\natexlab{b}}, \mnras, 444, 1425

\bibitem[{{Pijloo} {et~al.}(2015){Pijloo}, {Portegies Zwart}, {Alexander},
  {Gieles}, {Larsen}, {Groot}, \& {Devecchi}}]{pijlooetal15}
{Pijloo}, J.~T., {Portegies Zwart}, S.~F., {Alexander}, P.~E.~R., {et~al.}
  2015, \mnras, 453, 605

\bibitem[{{Rubele} {et~al.}(2011){Rubele}, {Girardi}, {Kozhurina-Platais},
  {Goudfrooij}, \& {Kerber}}]{rubeleetal2011}
{Rubele}, S., {Girardi}, L., {Kozhurina-Platais}, V., {Goudfrooij}, P., \&
  {Kerber}, L. 2011, \mnras, 414, 2204

\bibitem[{{Rubele} {et~al.}(2010){Rubele}, {Kerber}, \&
  {Girardi}}]{rubeleetal2010}
{Rubele}, S., {Kerber}, L., \& {Girardi}, L. 2010, \mnras, 403, 1156

\bibitem[{{Schmidt-Kaler}(1982)}]{shk82}
{Schmidt-Kaler}, T. 1982, {Landolt-B{\"o}rnstein: Numerical Data and Functional
  Relationships in Science and Technology, New Series, group VI, Vol. 2b}

\bibitem[{{Spitzer} \& {Hart}(1971)}]{sh71}
{Spitzer}, Jr., L. \& {Hart}, M.~H. 1971, \apj, 164, 399

\bibitem[{{Stetson} {et~al.}(1990){Stetson}, {Davis}, \& {Crabtree}}]{setal90}
{Stetson}, P.~B., {Davis}, L.~E., \& {Crabtree}, D.~R. 1990, in Astronomical
  Society of the Pacific Conference Series, Vol.~8, CCDs in astronomy, ed.
  G.~H. {Jacoby}, 289--304

\bibitem[{{Subramanian} \& {Subramaniam}(2009)}]{ss09}
{Subramanian}, S. \& {Subramaniam}, A. 2009, \aap, 496, 399

\bibitem[{{Subramanian} \& {Subramaniam}(2010)}]{ss10}
{Subramanian}, S. \& {Subramaniam}, A. 2010, \aap, 520, A24

\bibitem[{{van der Marel} \& {Kallivayalil}(2014)}]{vdmk14}
{van der Marel}, R.~P. \& {Kallivayalil}, N. 2014, \apj, 781, 121

\bibitem[{{Wilkinson} {et~al.}(2003){Wilkinson}, {Hurley}, {Mackey}, {Gilmore},
  \& {Tout}}]{wilkinsonetal03}
{Wilkinson}, M.~I., {Hurley}, J.~R., {Mackey}, A.~D., {Gilmore}, G.~F., \&
  {Tout}, C.~A. 2003, \mnras, 343, 1025

\end{thebibliography}

%to be uncommented before sending to editor

\end{document}